\documentclass[twocolumn,prb,superscriptaddress,showpacs,amsmath,amssymb]{revtex4}
\usepackage{bm,graphicx,dcolumn}
\newcommand{\piD}{\mathcal{D}}

\renewcommand{\Im}[1]{{\rm Im}\,#1}

\newcommand{\expec}[1]{\langle #1 \rangle}

\newcommand{\ket}[1]{|#1 \rangle}

\begin{document}
\title{Macroscopic quantum tunneling in multigap superconducting
Josephson junctions: Enhancement of escape rate via quantum fluctuations of
Josephson-Leggett mode }

\affiliation{
CCSE, Japan Atomic Energy Agency, 
6-9-3 Higashi-Ueno Taito-ku, Tokyo 110-0015, Japan}
\affiliation{
Institute for Materials Research, Tohoku University, 2-1-1 Katahira
Aoba-ku, Sendai 980-8577, Japan} 
\affiliation{
CREST(JST), 4-1-8 Honcho, Kawaguchi, Saitama 332-0012, Japan}
\affiliation{
JST, TRIP, 5 Sambancho Chiyoda-ku, Tokyo 102-0075, Japan}
\author{Yukihiro Ota}
\affiliation{
CCSE, Japan Atomic Energy Agency, 
6-9-3 Higashi-Ueno Taito-ku, Tokyo 110-0015, Japan}
\affiliation{
CREST(JST), 4-1-8 Honcho, Kawaguchi, Saitama 332-0012, Japan}
\author{Masahiko Machida}
\affiliation{
CCSE, Japan Atomic Energy Agency, 
6-9-3 Higashi-Ueno Taito-ku, Tokyo 110-0015, Japan}
\affiliation{
CREST(JST), 4-1-8 Honcho, Kawaguchi, Saitama 332-0012, Japan}
\affiliation{
JST, TRIP, 5 Sambancho Chiyoda-ku, Tokyo 102-0075, Japan}
\author{Tomio Koyama}
\affiliation{
Institute for Materials Research, Tohoku University, 
2-1-1 Katahira Aoba-ku, Sendai 980-8577, Japan}
\affiliation{
CREST(JST), 4-1-8 Honcho, Kawaguchi, Saitama 332-0012, Japan}
\date{\today}

\begin{abstract}
We theoretically study the macroscopic quantum tunneling (MQT) in a
 hetero Josephson junction formed by a conventional single-gap
 superconductor and a multigap superconductor such as and
 iron-based superconductors and $\mbox{MgB}_{2}$. 
In such a Josephson junction more than one phase difference is defined.  
We clarify their phase dynamics and construct a theory for the MQT in
 the multigap Josephson junctions.   
The dynamics of the phase differences are strongly affected by the
 Josephson-Leggett mode, which is the out-of-phase oscillation mode of
 the phase differences. 
The escape rate is calculated in terms of the effective action
 renormalized by the Josephson-Leggett mode at zero-temperature limit. 
We successfully predict drastic enhancement of the escape rate when the
 frequency of the Josephson-Leggett mode is less than the
 Josephson-plasma frequency. 
\end{abstract}

\pacs{74.50.+r}
\maketitle

Macroscopic quantum tunneling
(MQT)\,\cite{Calderira;Leggett:1981,Rajaraman:1987,Simanek:1994,Larkin;Varlamov:2005}
is a counterintuitive phenomenon in quantum mechanics appearing in a macroscopic level 
and has been observed in various fields of physics such as condensed matters, nuclei, 
cosmology, etc. 
This phenomenon has still attracted a great interest in physics communities. 
In particular, the MQT in Josephson junctions, which is observed in a
switching event at low
temperature\,\cite{Simanek:1994,Larkin;Varlamov:2005}, has been
intensively studied because it is promising for applications to a Josephson phase qubit
\,\cite{You;Nori:2005,Clarke;Wilhelm:2008,Nakahara;Ohmi:2008}. 

In this paper we explore the physics of MQT in an unexplored type of Josephson
junctions which has multiple tunneling channels. 
Such a Josephson junction can be fabricated by using recently 
discovered iron-based
superconductors\,\cite{Kamihara;Hosono:2008,Rotter;Johrendt:2008,Paglione;Greene:2010}
or $\mbox{MgB}_{2}$\,\cite{Nagamatsu;Akimutsu:2001,Brinkman;Rowell:2007,Xi:2008}, because these 
superconductors are multiband ones having more than one
disconnected Fermi surfaces and the superconducting gap can be
individually well defined on each Fermi surface. 
In a Josephson junction made of multigap superconductors one may expect that the
superconducting tunneling current has multiple channels between the two
superconducting electrodes \cite{Ota;Matsumoto:2009,Koyama;Machida:2010,Ota;Koyama:2010,Agterberg;Janko:2002}. 
We construct a theory for the quantum switching (i.e., MQT) in Josephson 
junctions with multiple tunneling channels. 
To our knowledge no theory has been formulated for the MQT in multigap systems. 
The theory predicts that the escape rate, i.e., the rate of quantum
tunneling, is drastically enhanced compared with that in conventional
single-channel systems. 

In multigap superconductors a collective mode called the Leggett's mode
\,\cite{Leggett:1966,Sharapov;Beck:2002,Ota;Aoki:2010} appears in the
low energy region, which is an out-of-phase oscillation mode of the
superconducting phases. 
In Ref.\,\cite{Ota;Matsumoto:2009} a theory for the Josephson effect in
superconducting hetero junctions made of a single-gap superconductor and
a two-gap one is formulated. 
In such Josephson junctions, because two kinds of gauge-invariant phase
differences can be defined, there are two phase oscillation modes, i.e.,
the in-phase mode and the out-of-phase one, which correspond,
respectively, to the Josephson-plasma and the Josephson-Leggett (JL) mode.  
In this paper we construct a theory for the MQT in superconducting
hetero junctions, incorporating the degree of freedom of the
JL mode into the quantum switching event from non-voltage
to voltage states.  
It is shown that the zero-point motion of the JL mode 
significantly enhances the MQT escape rate when its frequency is less than the Josephson-plasma 
frequency.  
We also point out that the ratio 
$E_{\rm J}/E_{\rm in}$ in addition to $E_{\rm J}/E_{\rm C}$ governs the boundary between the 
classical and quantum regimes, where $E_{\rm C}$, $E_{\rm J}$ and $E_{\rm in}$ are, respectively, 
the charging energy, the Josephson coupling energy between the two superconductors and the 
inter-band Josephson coupling energy in the two-gap superconductor. 

\begin{figure}[tbp]
\scalebox{0.5}[0.5]{\includegraphics{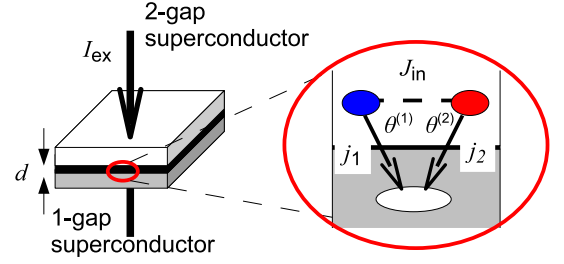}}
\caption{(Color online) Schematic view of a SIS hetero Josephson junction. We have two tunneling channels 
between the two superconductors with the critical current densities
 $j_{1}$ and $j_{2}$ as indicated in the right panel. In the upper
 two-gap electrode the inter-band Josephson coupling with the 
coupling constant $J_{\rm in}$ exists. }
\label{fig:junction}
\end{figure}

Consider a hetero Josephson junction made of a single-gap superconductor and a two-gap 
one\,\cite{Ota;Matsumoto:2009,Agterberg;Janko:2002,Ota;Koyama:2010}, as shown schematically 
in Fig.\,\ref{fig:junction}. Such a junction has been already fabricated, using the multigap 
superconductors $\mbox{MgB}_{2}$\,\cite{Shimakage;Tonouchi:2004} or iron-based 
superconductors\,\cite{Zhang;Takeuchi:2009,Schmidt;Holzapfel;2010}.
In this system one can define two gauge-invariant phase differences,  $\theta^{(1)}$ and
$\theta^{(2)}$. Then, the Josephson current density between the two superconducting 
electrodes is given by the sum of the superconducting currents in the two tunneling channels as   
\(
j_{1}\sin\theta^{(1)} + j_{2}\sin\theta^{(2)}
\), 
where $j_{i}$ is the Josephson critical current density in the $i$th tunneling channel. 
When a voltage $v$ appears between the two superconducting electrodes, the gauge-invariant 
phase differences show temporal evolution satisfying the generalized
Josephson relation\,\cite{Ota;Matsumoto:2009} 
\begin{equation}
 \frac{\alpha_{2}}{\alpha_{1}+\alpha_{2}}\dot{\theta}^{(1)}
+
 \frac{\alpha_{1}}{\alpha_{1}+\alpha_{2}}\dot{\theta}^{(2)}
= 
\frac{2e\Lambda}{\hbar}v,  
\label{eq:JR}
\end{equation}
with $\alpha_{i} =\epsilon \mu_{i}/d$ and 
$
 \Lambda=1+\alpha_{1}\alpha_{2}/(\alpha_{1}+\alpha_{2}), 
$
where $\epsilon$ is the dielectric constant of the insulator with a thickness $d$ and 
$\mu_{i}$ is the charge screening length due to the electrons in the $i$th band. 
The constant $\alpha_{i}$ is related to the charge compressibility in the two-gap 
superconducting electrode\,\cite{Ota;Koyama:2010}. 

As shown in Ref.\,\cite{Ota;Matsumoto:2009}, the Lagrangian 
in the hetero Josephson junction with a in-plane area $W$ and capacitance 
\(
C = \epsilon W/4\pi d
\)
is expressed as  
\begin{subequations}
\begin{eqnarray}
&&
L
=
\frac{1}{2}\frac{\hbar^{2}C}{(2e)^{2}}
\left(
\frac{\dot{\theta}^{2}}{\Lambda}
+
\frac{\dot{\psi}^{2}}{\alpha_{1}+\alpha_{2}}
\right) 
- V 
+ E_{\rm J}\frac{I_{\rm ex}}{I_{\rm c}}\theta, 
\label{eq:Lagrangian}\\
&&
V
=
-E_{\rm J1}\cos\theta^{(1)}
-E_{\rm J2}\cos\theta^{(2)}
-\kappa E_{\rm in}\cos\psi
\label{eq:Josephson_E},
\end{eqnarray}
\end{subequations} 
under a bias current $I_{\rm ex}$ in the absence of an external magnetic field, 
where $\theta$ and $\psi$ are the center-of-mass phase 
difference and the relative phase difference defined as 
\begin{equation*}
\theta 
= 
\frac{\alpha_{2}}{\alpha_{1}+\alpha_{2}}\theta^{(1)}
+
\frac{\alpha_{1}}{\alpha_{1}+\alpha_{2}}\theta^{(2)},
\quad
\psi 
= \theta^{(1)}-\theta^{(2)}.
\end{equation*}
The first two terms in Eq.\,(\ref{eq:Josephson_E}) are the Josephson
coupling energies with the coefficients 
\(
E_{{\rm J}i} = \hbar Wj_{i}/2e
\)
and the third term represents the inter-band coupling energy, where the
coefficient $E_{\rm in}$ is expressed as 
$E_{\rm in}=\hbar W|J_{\rm in}|/2e$\,\cite{Ota;Matsumoto:2009}.  
Since the ``inter-band current'' $J_{\rm in}$ can take both signs,
depending on the gap symmetry, we introduce the sign factor 
$\kappa = J_{\rm in}/|J_{\rm in}|$. 
The total critical Josephson current
$I_{\rm c}$ and the coefficient $E_{\rm J}$ in the last term in
Eq.\,(\ref{eq:Lagrangian}) are defined as 
$I_{\rm c}=W|j_{1}+\kappa j_{2}|$ and   
\(
E_{\rm J} = \hbar I_{\rm c}/2e
\). 
We note that the voltage $v$ appearing in the junction is related to only
$\theta$, as seen in Eq.\,(\ref{eq:JR}). 

From Eq.\,(\ref{eq:Lagrangian}) one can derive the Euler-Lagrange
equation for the center-of-mass  phase difference $\theta$ as 
\begin{equation}
 \Lambda^{-1}\ddot{\theta}
+
 \omega_{{\rm P}1}^{2}\sin\theta^{(1)}
+
 \omega_{{\rm P}2}^{2}\sin\theta^{(2)}
=
 \omega_{\rm P}^{2}\frac{I_{\rm ex}}{I_{\rm c}},
\label{eq:cl_in_eq}
\end{equation}
with 
\(
\hbar\omega_{{\rm P}i}
=
\sqrt{2 E_{\rm C}E_{{\rm J}i}}
\) and 
\(
\hbar\omega_{\rm P}
=
\sqrt{2 E_{\rm C}E_{\rm J}}
\). 
We note that $\omega_{{\rm P}i}$ is the Josephson-plasma
frequency in the $i$th tunneling channel. 
From Eq.\,(\ref{eq:Lagrangian}) we also have the  Euler-Lagrange
equation for $\psi$ as 
\begin{equation}
\ddot{\psi}
+\kappa\omega_{\rm JL}^{2}\sin\psi
=
-
\alpha_{1}\omega_{{\rm J}1}^{2}\sin\theta^{(1)}
+
\alpha_{2}\omega_{{\rm J}2}^{2}\sin\theta^{(2)},
\label{eq:cl_rlt_eq}
\end{equation}
where $\omega_{\rm JL}$ is the frequency of the JL
mode\,\cite{Ota;Matsumoto:2009} given as   
\(
\hbar\omega_{\rm JL} = \sqrt{2(\alpha_{1}+\alpha_{2})E_{\rm C}E_{\rm in}}
\). 
The above two equations are coupled since $\theta^{(1)}$ and $\theta^{(2)}$ are functions 
of $\theta$ and $\psi$. 
We note that the bias current is the source for the
time evolution of $\theta$ but not for $\psi$, which is consistent with
the generalized Josephson relation (\ref{eq:JR}). 
It should be also noted that we have two characteristic energy scales, the
Josephson-plasma frequency $\omega_{{\rm J}i}$ and the JL
one $\omega_{\rm JL}$, in this system. 

\begin{figure}[bp]
\centering
\scalebox{0.45}[0.45]{\includegraphics{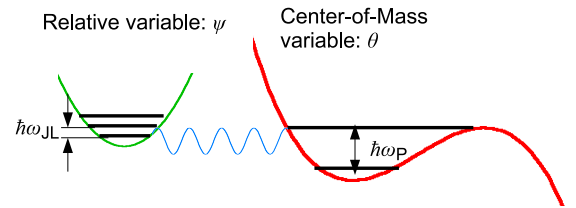}}
\caption{(Color online) Schematic  energy diagram for the fully ``quantized'' system with two quantum variables 
$\theta$ and $\psi$. In the case where $\psi$ is weakly oscillating within a potential well, the 
energy levels of $\psi$ coincide with those of a harmonic oscillator with frequency 
$\omega_{\rm JL}$. The energy levels of $\theta$ is corrected by the
 quantum oscillations of $\psi$. }
\label{fig:total_q_sys}
\end{figure}
\begin{figure}[htbp]
\scalebox{0.9}[0.9]{\includegraphics{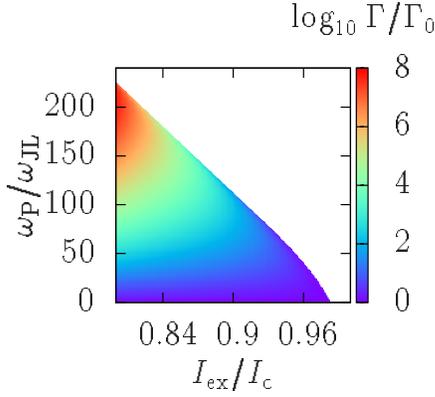}}
\caption{(Color online) Ratio  $\Gamma/\Gamma_{0}$ in the case of 
$E_{\rm J}/E_{\rm C}=10^{2}$. We assume that $\alpha_{1}=\alpha_{2}=0.1$ and
 $j_{1}=j_{2}$ for simplicity.}
\label{fig:ratio_mqt}
\end{figure}
\begin{figure}[htbp] 
\scalebox{0.45}[0.45]{\includegraphics{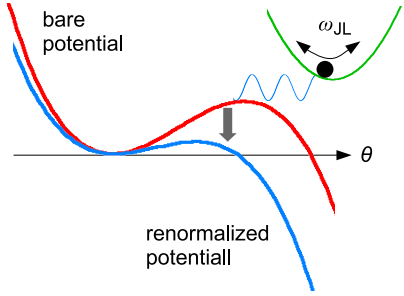}}
\caption{(Color online) Renormalized potential $V_{\rm cm,eff}$. The potential for $\theta$ is modified by the  
zero-point fluctuation of the JL mode.  }
\label{fig:renorm_pot}
\end{figure}

Let us now study the macroscopic quantum effects in the Josephson junction with multiple
tunneling channels on the basis of the Lagrangian (\ref{eq:Lagrangian}) and evaluate the MQT 
escape rate. In the following we assume $\kappa >0$, since the case of $\kappa <0$ shows 
qualitatively no difference. 

Suppose that the switching to the voltage state is induced by the
quantum tunneling of the phase differences $\theta^{(1)}$ and
$\theta^{(2)}$ which are confined inside a potential well. 
When both $\theta^{(1)}$ and $\theta^{(2)}$ show the tunneling at the
switching, its transition probability is given by the expectation value
of the time evolution operator with respect to the state
$\ket{\theta^{(1)}=0,\theta^{(2)}=0}(=\ket{\theta=0,\psi=0})$\,\cite{Simanek:1994}, 
which yields the formula for the MQT escape rate as     
\begin{equation}
 \Gamma = \frac{2}{\hbar\beta}\Im{K(\{0\},\{0\};\beta)}.
\label{eq:mqt}
\end{equation}
Here, the symbol $\{0\}$ means $(\theta,\psi)=(0,0)$, and  
$\beta$ is the inverse temperature, $\beta=1/k_{\rm B}T$. 
The propagator 
\(
K(X,X^{\prime};\beta)
\) in Eq.\,(\ref{eq:mqt}) is expressed in terms of the imaginary time path-integral  
\begin{equation*}
K(X,X^{\prime};\beta) 
=
\int_{X(0)=X^{\prime}}^{X(\hbar\beta)=X} 
\piD\theta\piD\psi \,
e^{-\int_{0}^{\hbar\beta} d\tau \, L^{\rm E}/\hbar },
\end{equation*}
where $X=(\theta,\psi)$ and $L^{\rm E}$ is the Euclidean version of the 
Lagrangian\,(\ref{eq:Lagrangian}). Let us assume that $\psi$ is confined in a small region 
around $\psi=0$ at the tunneling, which is justified when the inter-band
coupling is not so strong. 
In this case one can utilize the expansion with respect to $\psi$. 
Then, up to the order of $\psi^2$  the Euclidean Lagrangian $L^E$ is
approximated as 
\(
L^{\rm E} = L^{\rm E}_{\rm cm} + L^{\rm E}_{\rm rlt} + L^{\rm E}_{\rm int}
\), where
\begin{subequations}
\begin{eqnarray}
&&
L^{\rm E}_{\rm cm}
=
\frac{\hbar^{2}}{4E_{\rm C}}
\left(
\frac{d\theta}{d\tau}
\right)^{2}
-
E_{\rm J}
\left(
\cos\theta + \frac{I_{\rm ex}}{I_{\rm c}}\theta
\right) , 
\label{eq:imL_cm}
\\
&&
L^{\rm E}_{\rm rlt}
=
\frac{\hbar^{2}}
{4 (\alpha_{1}+\alpha_{2}) E_{\rm C}}
\left(
\frac{d\psi}{d\tau}
\right)^{2}
+
\frac{1}{2}E_{\rm in}\psi^{2}, 
\label{eq:imL_rlt}\\
&&
L^{\rm E}_{\rm int}
=
g_{+} E_{\rm J}\psi^{2}\cos\theta
-
g_{-} E_{\rm J} \psi \sin\theta 
\label{eq:imL_int}.
\end{eqnarray}
\end{subequations}
Here, $E_{\rm J} = E_{{\rm J}1} + E_{{\rm J}2}$ and $\Lambda \approx 1$ is assumed. 
The coupling constants $g_{+}$ and $g_{-}$ in Eq.\,(\ref{eq:imL_int}) are defined as  
\(
g_{+}
=
(E_{{\rm J}1}/2E_{\rm J})
[\alpha_{1}/(\alpha_{1}+\alpha_{2})]^{2}
+
(E_{{\rm J}2}/2E_{\rm J})
[\alpha_{2}/(\alpha_{1}+\alpha_{2})]^{2}
\) and 
\(
g_{-}
=
(E_{{\rm J}1}/E_{\rm J})
[\alpha_{1}/(\alpha_{1}+\alpha_{2})]
-
(E_{{\rm J}2}/E_{\rm J})
[\alpha_{2}/(\alpha_{1}+\alpha_{2})]
\). 
We note that in the fully quantum case we have the discrete energy levels as schematically 
illustrated in Fig.\,\ref{fig:total_q_sys}.  
To calculate the escape rate $\Gamma$ in Eq.\,(\ref{eq:mqt}) we employ the mean field
approximation for $\psi$, that is, $\psi^{2}$ and $\psi$ in
Eq.\,(\ref{eq:imL_int}) are approximated with their expectation values. 
Then, at zero temperature we find $\expec{\psi}_{\rm th}=0$ and  
\begin{equation*}
 \expec{\psi^{2}}_{\rm th}(T=0)
=
\frac{\hbar}{2m_{\rm rlt}\omega_{\rm JL}},
\quad
m_{\rm rlt}
=
\frac{\hbar^{2}}{2(\alpha_{1}+\alpha_{2})E_{\rm C}}. 
\end{equation*}
The finite value of  $\expec{\psi^{2}}$ originates from the zero-point motion  of the 
``quantized'' JL mode. 
Under this approximation we find the effective 
Lagrangian of single degree of freedom as follows, 
\begin{subequations}
\begin{equation}
 L^{\rm E}_{\rm cm,eff}
=
\frac{\hbar^{2}}{4E_{\rm C}}\left(
\frac{d\theta}{d\tau}
\right)^{2}
+
V_{\rm cm,eff},
\label{eq:Lcm_renom}
\end{equation}
where $V_{\rm cm,eff}$ is the renormalized potential 
\begin{eqnarray}
&&
 V_{\rm cm,eff}
=
-E_{\rm J}\left[
(1-\varepsilon)\cos\theta + \frac{I_{\rm ex}}{I_{\rm c}}\theta
\right], 
\label{eq:eff_cm_pot}\\
&&
\varepsilon
=
g_{+}\expec{\psi^{2}}_{\rm th}
\approx
\frac{g_{+}}{\sqrt{2}}(\alpha_{1}+\alpha_{2})
\frac{\omega_{\rm P}}{\omega_{\rm JL}}\sqrt{
\frac{E_{\rm C}}{E_{\rm J}}
}.
\label{eq:rlt_zero_f}
\end{eqnarray}
\end{subequations} 
Then, in this approximation the expectation value $K(\{0\},\{0\};\beta)$
in Eq.\,(\ref{eq:mqt}) is reduced to 
\(
K(\{0\},\{0\};\beta=\infty) 
\approx 
\int_{\theta(0)=0}^{\theta(\hbar\beta=\infty)=0} 
\piD\theta \, 
\exp(-\hbar^{-1}\int_{0}^{\hbar\beta=\infty}L^{\rm E}_{\rm cm,eff} d\tau)
\), 
which can be evaluated in the standard instanton approximation\,\cite{Simanek:1994}. 
Hence, the MQT escape rate corrected by the zero-point motion of $\psi$ is 
\begin{equation}
\Gamma
=
12\omega_{\rm P}(I)
\sqrt{
\frac{3V_{0}}{2\pi\hbar\omega_{\rm P}(I)}
} 
\exp\left(
-\frac{36V_{0}}{5\hbar \omega_{\rm P}(I)}
\right),
\label{eq:mqt_with_rlt}
\end{equation}
where
\(
\omega_{\rm P}(I)
=
\omega_{\rm P}
[(1-\varepsilon)^{2}-I^{2}]^{1/4}
\), 
\(
V_{0}=\hbar^{2}[\omega_{\rm P}(I)]^{2}\cot^{2}\theta_{0}/3E_{\rm C}
\), 
\(
(1-\varepsilon)\sin\theta_{0}=I
\), and 
\(
I=I_{\rm ex}/I_{\rm c}
\). 

Figure \ref{fig:ratio_mqt} shows a contour map of the ratio $\Gamma/\Gamma_{0}$ 
in the ($I_{\rm ex}/I_{\rm c}$ vs. $\omega_{\rm P}/\omega_{\rm JL}$)-plane with 
$\Gamma_0$ being the escape rate without correction, i.e., $\varepsilon=0$. 
It is seen that the escape rate is drastically enhanced in a wide parameter region.   
In particular, the enhancement is pronounced in the region of large
$\omega_{\rm P}/\omega_{\rm JL}$. 
As seen in Eqs.\,(\ref{eq:Lcm_renom}) and (\ref{eq:eff_cm_pot}), the
Josephson coupling energy is renormalized by the zero point motion of
$\psi$ and the renormalized one is decreased from the bare one since
$\varepsilon >0$. 
As a result, the tunneling barrier for $\theta$ is lowered 
as schematically shown in Fig.\,\ref{fig:renorm_pot}, which causes the strong 
enhancement of the escape rate. 
In fact, $R(\varepsilon)\equiv V_{0}/\hbar\omega_{\rm P}(I)$ is smaller than 
$R(\varepsilon=0)$ for fixed $I$ when $0<\varepsilon <1$, that is, the exponent 
in Eq.\,(\ref{eq:mqt_with_rlt}) is decreased. Thus, the renormalization increases $\Gamma$. 
It should be also noted that the zero-point fluctuation becomes larger as the frequency
of the JL mode decreases. 
Thus, the considerable enhancement of $\Gamma$ occurs for the system with a 
lower value of $\omega_{\rm JL}$. 
The MQT in the conventional systems is subject to 
the ratio $E_{\rm J}/E_{\rm C}$, which is an important parameter for designing 
a superconducting Josephson qubit\,\cite{You;Nori:2005}.  
In the system with multiple tunneling channels the ratio 
$\omega_{\rm P}/\omega_{\rm JL}(\propto E_{J}/E_{\rm in})$ also affects the characteristics 
of the MQT. 

In this paper we have focused on the tunneling process,   
\(
\ket{\theta=0,\psi=0} \, \to \ket{0,0}
\) 
and clarified the effect of the JL mode on the MQT. 
We mention that such a process is not the unique one that contributes to the MQT rate 
in this system, since a system with two degrees of freedom generally have many tunneling 
routes. For example, the tunneling process in which the quantum switching in the $\theta^{(1)}$ 
channel takes place successively after the switching in the $\theta^{(2)}$ channel will be 
also possible in the present system\,\cite{Chizaki;Kawabata:2010}.  In this case the escape 
rate can be calculated from the transition process
$\ket{\theta^{(1)}=0}\,\to \ket{0}$ with $\theta^{(2)}=f(t)$, where
$f(t)$ is a time-dependent c-number function.  
This tunneling process is analogous to the MQT under a periodically time-dependent
perturbation\,\cite{Fisher:1988}. 
It is also noted that the relative phase difference $\psi$ 
might play a role of an environmental variable for $\theta$ through the term linear in $\psi$ 
in Eq.\,(\ref{eq:imL_int}). 
The MQT rate in this process can be evaluated, using the influential functional
integral method\,\cite{Calderira;Leggett:1981,Kawabata;Kato:2009}. 
The competition between 
the zero-point fluctuation and the ``dissipation'' occurs in this case. 
The enhancement via the JL mode may be superior to the reduction from such dissipation when
$g_{+}>|g_{-}|$. 

We also mention that our theory for the MQT in the hetero Josephson junctions can be 
extended to the case of intrinsic Josephson junctions (IJJ's) with
multiple tunneling channels\,\cite{Koyama;Machida:2010}.  
The MQT in such systems will be observed in several highly-anisotropic layered iron-based
superconductors recently
discovered\,\cite{Ogino;Shimoyama:2009,Nakamura;Hamada:2009,Kashiwaya;Kashiwaya:2010}.  
In the IJJ's correction due to the JL mode for the corporative MQT among
the junctions\,\cite{Machida;Koyama:2007,Savelev;Nori:2008} will be expected. 

Finally, we remark that the present theoretical prediction relies on the
coexistence of the Josephson-plasma and the JL modes. 
Since observation of Leggett's mode in a bulk $\mbox{MgB}_{2}$ sample has
been reported\,\cite{Blumberg;Karpinski:2007} and in junctions with
$\mbox{MgB}_{2}$\,\cite{Ponomarev;Bulychevd:2004,Brinkman;Moodera:2006},
we expect that such a collective mode can  
be detected in a junction system and the theory will be verified
experimentally. 

In summary, we have constructed a theory of the MQT in hetero Josephson junctions with
multiple tunneling channels. We have clarified that the zero-point fluctuation of the relative
phase differences brings about the drastic enhancement of the MQT escape rate. 
The enhancement is gigantic when the JL mode has a lighter mass than that of the 
Josephson-plasma.

YO thanks M. Nakahara, S. Kawabata, and Y. Chizaki for illuminating discussions.
TK was partially supported by Grant-in-Aid for Scientific Research (C)
(No. 22540358) from the Japan Society for the Promotion of Science.

\end{document}